\documentclass[a4paper]{jpconf}
\usepackage{graphicx}
\def\mWeq{M_{W}}
\def\mW{$\mWeq$}
\def\W{$W$}
\def\Z{$Z$}

\def\mt{$M_{T}$}

\def\gW{$\Gamma_{W}$}

\def\mTeq{M_{T}}
\def\mT{$\mTeq$}
\def\pt{$p_{T}$}
\def\ptleq{p_{T}^{\ell}}
\def\ptl{$\ptleq$}
\def\ptlv{$\vec{\ptleq}$}
\def\u{$\vec{U}$}

\def\Wenu{$W\rightarrow e\nu$}
\def\Wmunu{$W\rightarrow\mu\nu$}
\def\Wlnu{$W\rightarrow\ell\nu$}

\def\Zll{$Z\rightarrow\ell\ell$}

\def\metEq{{E}\!\!\!/_T }
\def\met{$\metEq$}
\def\ipb{$\rm pb^{-1}$}
\begin{document}
\title{W production and properties at CDF}

\author{Emily Nurse (for the CDF collaboration)}

\address{University College London, Gower Street, London, WC1E~6BT, UK}

\ead{nurse@fnal.gov}

\begin{abstract}
I present measurements of the \W\ boson charge asymmetry and the \W\ boson width (\gW) using 350~\ipb\ of CDF Run II data.
The charge asymmetry is the first direct measurement, which uses
a new technique to reconstruct the \W\ rapidity by constraining its mass; the result will further constrain
Parton Distribution Functions in future fits.
The width measurement relies on a fit to the \W\ transverse mass distribution.
We measure \gW\ = 2032 $\pm$ 71~MeV which is in good agreement with the Standard Model prediction.
\end{abstract}

\section{Introduction}
\W\ bosons are produced at the Tevatron proton-antiproton collider via quark-antiquark annihilation. 
They quickly decay into two fermions which, due to the large mass of the \W,
 typically have a high momentum in the direction transverse to the beam, \pt.
The leptonic decay channels have a clean experimental signature and can therefore
be utilised to make precision measurements of \W\ properties, such as its width (\gW) and mass (\mW), and
to probe the quantum-chromodynamic (QCD) aspects of the production mechanism.
This document describes a measurement of the \W\ charge asymmetry made in the $e\nu$ decay channel and \gW, made
in the $e\nu$ and $\mu\nu$ decay channels, using 350~\ipb\ of CDF Run II data.
\section{\W\ charge asymmetry}
Parton Distribution Functions (PDFs) are parameterised functional forms that describe the momentum distribution of partons in hadrons.
The parameters are constrained by many data sets from different experiments. The $u(\bar{u})$ quark carries on average a higher
fraction of the (anti)proton's momentum than the $d(\bar{d})$ quark, meaning that a $W^+(W^-)$ produced  via $u\bar{d}(d\bar{u})$ 
annihilation will tend to be boosted in the direction of the (anti)proton beam.
This results in a \W\ charge asymmetry, defined as:
\begin{equation}
A(y_{W}) = \frac{d\sigma(W^{+})/dy_{W} - d\sigma(W^{-})/dy_{W}}{d\sigma(W^{+})/dy_{W} + d\sigma(W^{-})/dy_{W}},
\end{equation}
where $y_{W}$ is the rapidity of the \W\ and $\sigma(W^{\pm})$ is the cross-section of \W\ production.
This distribution is sensitive to the difference between the $u$ and $d$ PDFs at the scale $Q^2 \approx M_{W}^{2}$
and is an important input to the global PDF fits. In particular the measurement will help to reduce the PDF uncertainty
on future \mW\ and \gW\ measurements.

\begin{figure}[h]
\includegraphics[width=16.5pc]{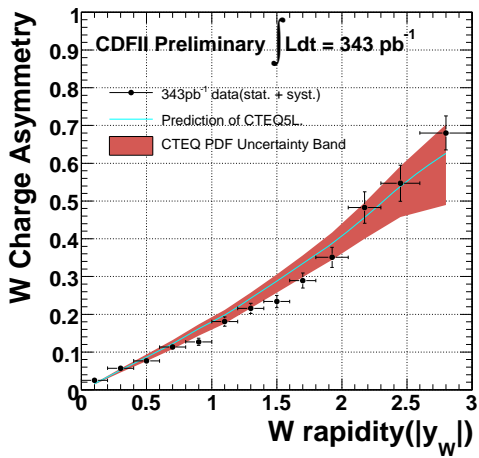}\hspace{2pc}%
\begin{minipage}[b]{17pc}\caption{\label{fig:asym}The \W\ charge asymmetry distribution
 compared to the CTEQ5L PDF prediction with the CTEQ6.1 error sets shown as the uncertainty band.}
\end{minipage}
\end{figure}

Since the neutrino momentum in the direction of the beam, $p_{z}^{\nu}$, is not known, traditionally a measurement of the 
electron charge asymmetry is made. This distribution is a convolution of $A(y_{W})$ and the \Wenu\ angular decay structure
and places a less strong constraint on PDFs than $A(y_{W})$.
This analysis extracts $A(y_{W})$ by constraining $M_{W} = 80.4~\rm GeV/c^2$, giving two possible solutions
for $p_{z}^{\nu}$. Each solution receives a probability weight according to the  decay structure 
and $\sigma(W^{\pm})$. The \W\ charge asymmetry is then extracted after correcting for detector effects.
The process is iterated to eliminate any dependence of the weighting factor on the asymmetry itself.
The measured \W\ charge asymmetry is compared to the prediction from PDFs using CTEQ5L and the 
corresponding error from the CTEQ6.1~\cite{cteq} error sets in Figure~\ref{fig:asym}. 
The experimental uncertainties are lower than the uncertainties from the PDFs indicating that they will help
to constrain the PDFs in future fits.
\section{W width measurement}
Since neutrinos are not detected in CDF, the invariant mass of the \W\ decay products cannot be reconstructed. 
Instead we reconstruct the transverse mass, which is defined as  $\mTeq = \sqrt{2\ptleq p_{T}^{\nu}(1 - \cos \phi_{\ell\nu})}$,
where \ptl\ and  $p_{T}^{\nu}$ are the \pt\ of the charged lepton and the neutrino respectively
and $\phi_{\ell\nu}$ is the azimuthal angle between them.
$p_{T}^{\nu}$ is inferred from the transverse momentum imbalance in the event.

A fast, parameterised Monte Carlo simulation is used to predict the \mT\ distribution as a function of \gW.
After adding background distributions to the Monte Carlo, a binned maximum-likelihood fit to the data is performed in the region 90--200~GeV/$c^2$ to extract \gW.
The fit is performed in the high \mT\ tail region, which is sensitive to the Breit-Wigner line-shape but less
sensitive to the Gaussian detector resolutions than the peak region.
These line-shape predictions depend on a number of production and detector effects, described below.

The \W\ \pt\ spectrum, which is non-zero due to initial state gluon radiation from the incoming quarks, 
is modelled with RESBOS~\cite{resbos}. QED corrections for up to one photon emission are simulated with 
Berends and Kleiss~\cite{bk}.  Systematic uncertainties arise from non-perturbative QCD parameters
affecting the \W\ \pt\ spectrum, and also from neglecting to model the effect of a second photon from the final state charged lepton.
PDFs affect the acceptance and kinematics of decay products. The Monte Carlo is
interfaced with the CTEQ6.1~\cite{cteq} PDFs, and their error sets are used to estimate the PDF uncertainty.

The simulation has a fast detector response model which is tuned to \Zll\ and \Wlnu\ data, systematic uncertainties come
predominantly from the limited statistics in the data.
The scale and resolution of the drift chamber muon momentum
measurement are calibrated using a fit to the $Z \rightarrow\mu\mu$ resonance peak.
The electron energy is measured in the calorimeter and the momentum is measured in the drift
chamber~\footnote{Since the mass of the electron is negligible, the true momentum and energy values are the same. 
                  However collinear photon radiation from the electron, which is clustered back into the calorimeter energy 
                  measurement, can decrease the track momentum measurement. These effects are well modelled in the simulation.}.
The well calibrated momentum measurement can therefore be used to calibrate both the calorimeter scale and resolution, using
the ratio between the electron energy measured in the calorimeter and the track momentum (E/p)
in \Wenu\ events. The scale and resolution are also obtained from the $Z \rightarrow ee$ resonance peak and the results
are combined.

The neutrino \pt\ is determined from the missing transverse energy, \met, in the detector. 
A recoil vector, \u, is defined as the vector sum of the transverse energy in all calorimeter towers, excluding those surrounding the lepton. 
The \met\ is then defined as -(\u\ + \ptlv).
The recoil is represented by a parameterised model, which is tuned to \Zll\ events.

The dominant systematic uncertainties on \gW\ come from modelling the lepton resolutions and the recoil as well as modelling
the background \mt\ distributions and normalisations.

\begin{figure}[h]
\begin{minipage}{16.5pc}
\includegraphics[width=16.5pc]{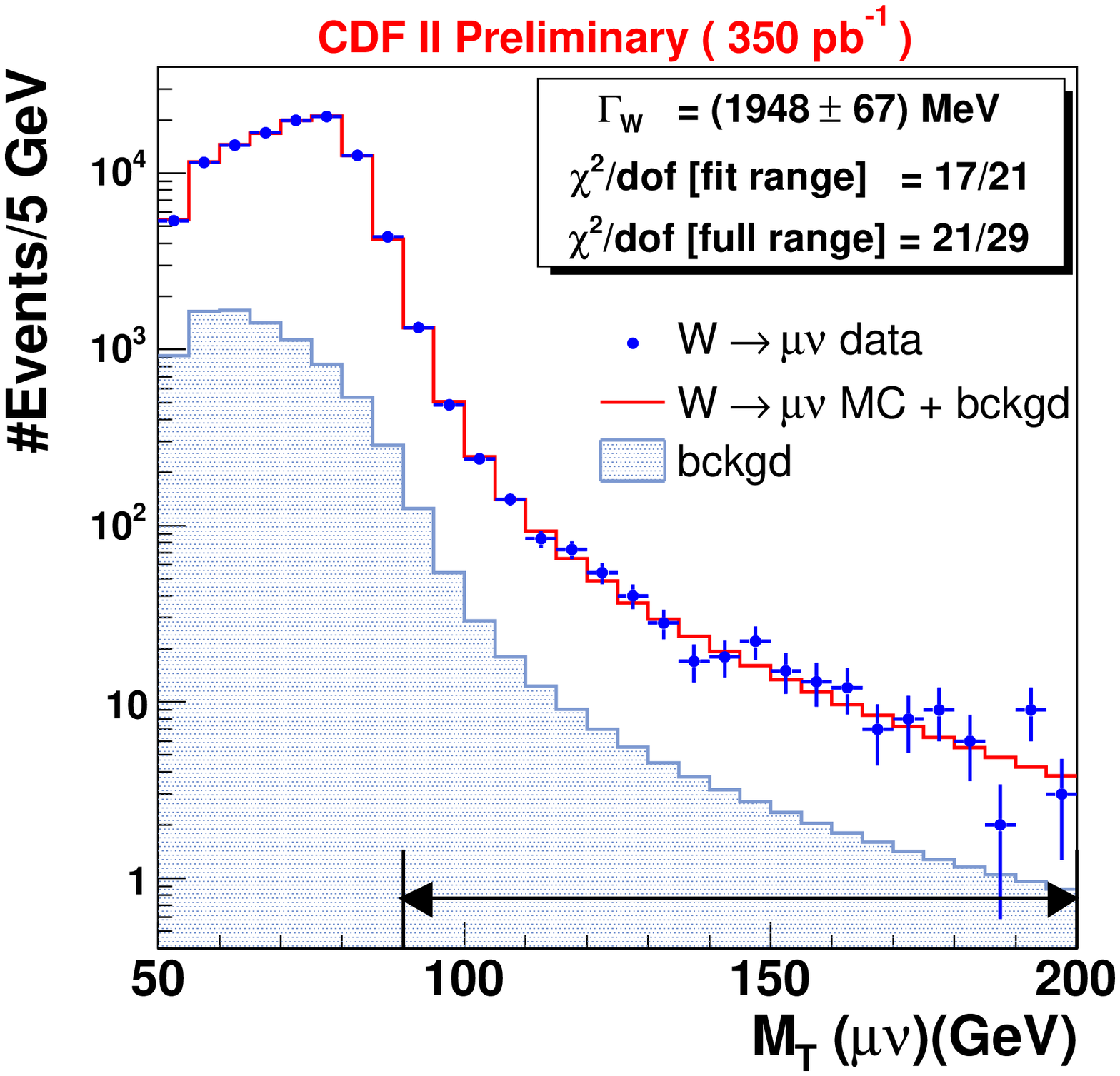}
\end{minipage}\hspace{2pc}%
\begin{minipage}{16.5pc}
\includegraphics[width=16.5pc]{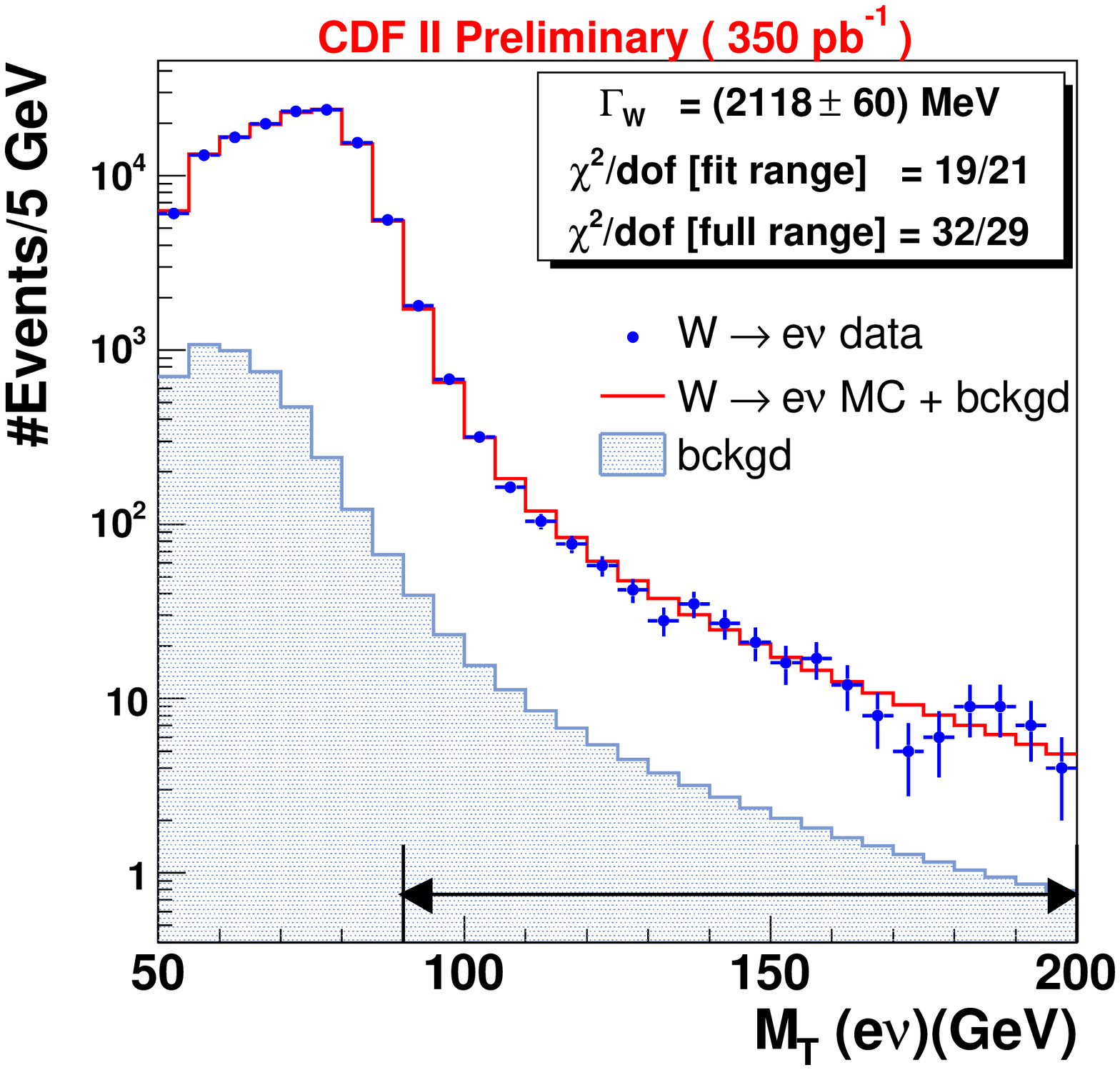}
\end{minipage} 
\caption{\label{fig:width-fits}Transverse mass fits for \gW\ in \Wmunu\ (left) and \Wenu\ (right) events. The fits are performed in the region 90--200~GeV.}
\end{figure}

Figure~\ref{fig:width-fits} shows the  \mT\ fits for \gW\ in the  muon and electron decay channels. The results are combined to give the final result 
\gW\ = 2032 $\pm$ 71~MeV, the world's most precise single measurement, which is in good agreement with the Standard Model prediction of 2093 $\pm$ 3~MeV~\cite{pdg}. This result reduces the
world average central value by 44~MeV and uncertainty by 22\%. CDF has also made an indirect measurement of \gW, 
obtained from a measurement of the ratio of the cross-section times branching ratio for leptonically decaying \W's and \Z's~\cite{indirect}.
This measurement yields \gW~=~2092~$\pm$~42~MeV, which is consistent with the direct measurement.

\end{document}